\documentclass[onecolumn,%superscriptaddress,showpacs,
amsmath,amssymb,footinbib]{revtex4}
\usepackage{graphicx}% Include figure files
\usepackage{subfigure}
\usepackage{amsthm}
\usepackage{color}
\usepackage[all]{xy}
\usepackage{tikz}
\usepackage{dsfont}
\usetikzlibrary{positioning}
\usepackage{braket}

\usepackage{appendix}

\newtheorem{Thm}{Theorem}
\newtheorem{Lem}[Thm]{Lemma}
\newtheorem{Prop}[Thm]{Proposition}

\newtheorem{Def}[Thm]{Definition}
\newtheorem{Rem}[Thm]{Remark}
\newtheorem{algorithm}[Thm]{Algorithm}
\newtheorem{Claim}{Claim}

\newcommand{\inn}[2]{\left\langle{#1}|{#2}\right\rangle}

\begin{document}

\title{Quantum algorithm based on the $\varepsilon$-random linear disequations for the continuous hidden shift problem 
}

\author{Eunok Bae}\email{eobae@khu.ac.kr}
\affiliation{
Department of Mathematics and Research Institute for Basic Sciences,
Kyung Hee University, Seoul 02447, Korea}

\author{Soojoon Lee}\email{level@khu.ac.kr}
\affiliation{
Department of Mathematics and Research Institute for Basic Sciences,
Kyung Hee University, Seoul 02447, Korea}

\begin{abstract}
There have been several research works on the hidden shift problem, quantum algorithms for the problem, 
 and their applications. However, all the results have focused on discrete groups with discrete oracle functions. 
 In this paper, we define the continuous hidden shift problem on $\mathbb{R}^n$ with a continuous oracle function as an extension of the hidden shift problem, 
 and also define the $\varepsilon$-random linear disequations which is a generalization of the random linear disequations.
 By employing the newly defined concepts, 
 we show that there exists a quantum computational algorithm which solves this problem in time polynomial in $n$.
%\keywords{Quantum algorithm \and Continuous hidden shift problem \and $\varepsilon$-random linear disequations}
%\PACS{03.67.Dd}
%\subclass{68Q12, 81P68}
\end{abstract}

\maketitle

%-----------------------------------------%
%            Introduction                 %
%-----------------------------------------%

\section{Introduction}
\label{Introduction}

 Quantum computers can solve certain problems exponentially faster than classical computers 
 by taking advantage of the quantum mechanical properties such as quantum interference and superposition. 
 Many researchers have been studying algebraic problems which can be solved more efficiently on a quantum computer than a classical computer, for instance, hidden subgroup problem~\cite{Simon,Shor,EH00,EHK04,Kup05,Hal05,SV05,Kup13}, hidden shift problem~\cite{FIM+03,DHI03,CD07,Ivanyos}, hidden polynomial problem~\cite{CW07,DDW07,DHIS14}, and hidden symmetry subgroup problem~\cite{DISW13,KBL13}.
 In particular, the hidden shift problem has provided a framework to solve various problems 
such as the shift Legendre symbol problem~\cite{DHI03}, 
Gauss sum estimation~\cite{DS02}, and 
the stabilizer problem~\cite{FIM+03}.
It has been shown that several interesting and important problems have been related to the hidden shift problem. 
For example, it was proved that 
the hidden shift problem for the abelian group $\mathbb{Z}_N$ can be used to solve 
some lattice problem over $\mathbb{Z}_N$~\cite{EH00,Regev}, and 
it was also discovered that an efficient algorithm of the hidden shift problem for the symmetric group $S_n$ 
would yield an efficient algorithm for the graph isomorphic problem~\cite{CW07}.
 
 The hidden shift problem can be cast in the following terms: 
 Let $f_0$ and $f_1$ be two injective functions from a finite group $G$ to a finite set satisfying that
 there exists an element $u$ in $G$ such that the equality $f_0(x)=f_1(xu)$ holds for all $x$ in $G$.
 The task is to find the \emph{hidden shift} $u$. 
 Although there is no general algorithm to solve the hidden shift problem even for abelian groups,
 it has been known that there are efficient quantum algorithms to solve the problem for some groups while no classical algorithm for solving this problem in time $O(\text{poly}(n))$ is known~\cite{DHI03,GRR,Roetteler16,Roetteler09}.
Friedl {\it et al.}~\cite{FIM+03} found an efficient quantum algorithm 
 for the hidden shift problem over $\mathbb{Z}_p^n$ for any fixed prime number $p$, and 
 a similar work for the problem over the group $\mathbb{Z}_{p^k}^n$ has been done by Ivanyos~\cite{Ivanyos},
 where $p^k$ is any fixed prime power.
The hidden shift problem over $\mathbb{Z}_{p^k}^n$ can be solved in time polynomial in $n$ with a small error by using a quantum computer.
 However, when $m$ is not a prime power, the hidden shift problem for the group $\mathbb{Z}_{m}^n$ still remains unsolved. 

  All known results on the hidden shift problem have been concerned with only discrete groups with discrete oracle functions.
 Thus, it is natural to ask 
 whether there exists an efficient quantum algorithm for solving the continuous hidden shift problem,
 which is the hidden shift problem on a group with a continuous oracle function. 
 Considering a continuous version of an algebraic  problem can be helpful
 to solve unsolved problems as in the results of Eisentr\"{a}ger {\it et al.}~\cite{EHKS14}.
 They found an efficient quantum algorithm for solving a continuous hidden subgroup problem on $\mathbb{R}^n$ with a continuous oracle function hiding a hidden subgroup
 to compute the unit group of an arbitrary degree number field.
 It was also shown that the algorithm can pose a threat to certain lattice-based cryptosystems~\cite{EHKS14,BS16}.
 
 In this paper, we present the continuous hidden shift problem for $\mathbb{R}^n$ with a continuous oracle function hiding a hidden shift. 
 To deal with the continuous inputs, we truncate the domain using a large enough number $2^q$ and discretize the inputs coordinatewise. So, the oracle inputs considered as the elements in $\mathbb{Z}_{2^q}^n$.
Remark that the author in Ref.~\cite{Ivanyos} had used the random linear disequations to solve the hidden shift problem over $\mathbb{Z}_{p^k}^n$. However, we cannot apply the same method directly to the discretized inputs of the continuous hidden shift problem since we differently use a continuous oracle function instead of a discrete oracle function. 
Thus, we newly define the $\varepsilon$-random linear disequations which is a generalization of the random linear disequations, 
and we construct a quantum algorithm for solving the continuous hidden shift problem on $\mathbb{R}^n$ by employing the method.
 
 Our paper is organized as follows. 
 In Sec.~\ref{sec:ctsHTP}, we give the definition of the continuous hidden shift problem on $\mathbb{R}^n$, 
 and introduce our main result.
 In Sec.~\ref{sec:eRLD}, we define two types of the $\varepsilon$-random linear disequations problem, 
 which are the search version and the decision version, 
 and show that we can efficiently solve the decision version of this problem.
 In Sec.~\ref{sec:algorithm}, we present our algorithm to solve the continuous hidden shift problem on $\mathbb{R}^n$ by reducing the problem to the decision version of the $\varepsilon$-random linear disequations problem on $\mathbb{R}^n$.
 In Sec.~\ref{sec:analysis}, we analyze the efficiency of our algorithm, and 
 conclude with discussion on our results in Sec.~\ref{sec:discussion}.

%-------------------------------------------------------%
%       Hidden shift problem on a continuous group      %
%-------------------------------------------------------%

\section{Continuous hidden shift problem on $\mathbb{R}^n$}
\label{sec:ctsHTP}

To deal with the continuous hidden shift problem on $\mathbb{R}^n$, 
we need a suitable definition. 
The following definition can be considered as a continuous version of the original hidden shift problem.
 
 \begin{Def}[Continuous hidden shift problem over $\mathbb{R}^n$]
\label{def:cHSP}
 Let $S$ be the set of unit vectors in a Hilbert space $\mathcal{H}$. 
For two injective functions $f_0$ and $f_1$ from $\mathbb{R}^n$ to $S$,
let $f: \mathbb{R}^n \times \mathbb{Z}_2 \rightarrow S$ be defined by $f(x,a):=f_a(x)$
 with the following promises:

\begin{enumerate}
\item $f(x,0)=f(x+u,1)$ 
for all $x \in \mathbb{R}^n$ and for some $u \in \mathbb{R}^n$.
\item There exists $\alpha>0$ such that $\left\Arrowvert \ket{f(x,a)}- \ket{f(y,b)}\right\Arrowvert_\mathcal{H} \leq \alpha \cdot \left\Arrowvert x-y-(a-b)u \right\Arrowvert $ 
for all $x,y \in \mathbb{R}^n$ and $a,b \in \mathbb{Z}_2$,
where $\ket{f(\cdot, \cdot)}$ is a real coefficient pure state corresponding to $f(\cdot, \cdot)$,
$\left\Arrowvert\cdot\right\Arrowvert_\mathcal{H}$ is the norm 
induced by the inner product $\inn{\cdot}{\cdot}$ on $\mathcal{H}$, 
and $\left\Arrowvert\cdot\right\Arrowvert$ is the Euclidean norm on $\mathbb{R}^n$.

\item\label{condition3} If $\left\Arrowvert x-y-(a-b)u \right\Arrowvert \geq r$, then  
$\left |\inn{f(x,a)}{f(y,b)} \right| \le \xi$.
\end{enumerate}
The continuous hidden shift problem over $\mathbb{R}^n$ with positive real parameters $(\alpha,r,\xi)$ is to find an $\eta$-approximation $u_a$ of the hidden shift $u$ such that $\left\Arrowvert u_a - u \right\Arrowvert < \eta$.
\end{Def}
 
 Note that %the given oracle function $f$ is efficiently computable, and 
 the positive constant $\alpha$ in the condition~2 of Definition~\ref{def:cHSP} is called a \emph{Lipschitz constant} of the function $f$. We can reformulate the condition~2 by using the inner product instead of the norm as follows:
\begin{enumerate}\setcounter{enumi}{1}
\item There exists $\alpha>0$ such that $1- \inn{f(x,a)}{f(y,b)} \le \frac{\alpha^2}{2} \left\Arrowvert x-y-(a-b)u \right\Arrowvert^2$ for all $x,y \in \mathbb{R}^n$ and $a,b \in \mathbb{Z}_2$.
\end{enumerate}

The algorithm to find an approximation of $u$ with high probability in time polynomial in $n$ should require certain conditions of the parameters $\alpha, r$, and $\xi$ with respect to an assumption of the hidden shift $u$ which is not too large. The detailed statement of our main theorem is as follows.

\begin{Thm}
\label{thm:main}
Suppose that $u=(u_1, u_2, \ldots, u_n) \in \mathbb{R}^n$ satisfies $u_i^2 \leq \sqrt{2^q}$ for all $i$ and for some positive integer $q$.
Let $\delta = \frac{1}{\sqrt{2^{q}}}$. Then there exists a quantum algorithm to find the $\frac{\delta}{2}$-approximation $\delta\tilde{u}$ of the hidden shift $u$ in $\mathbb{R}^n$ in time polynomial in $n$ with a small constant error if the parameters $(\alpha, r, \xi)$ of the oracle function are chosen as $\alpha^2 < 2^{q+3}$, $r = \frac{\delta}{2}$, and $\xi$ is less than both 
\begin{equation}
     \frac{1}{2^{qn+1}}
     \left(3
     -
     %+
     \left(1-2^{-\frac{q}{4}}\right)^n 
     %A 
     %\left(
     %1-\frac{\alpha^2}{2^{q+3}}
     %\right)
     %^{\frac{|A|+A}{2A}}
     \right)
    \label{xi_1}
 \end{equation}
 and
 \begin{equation}
     \frac{1}{3 \cdot 2^{qn}}
     \left(
    \left(1-2^{-\frac{q}{4}}\right)^n
    \left( 1-\frac{\alpha^2}{2^{q+3}}
    \right)
    - A_0
    \left(
    \left(1-2^{-\frac{q}{4}}\right)^n
    \left( 1-\frac{\alpha^2}{2^{q+3}}
    \right)
    \right)^{\frac{|A_0|+A_0}{2A_0}}
    \right),
    \label{xi_2}
 \end{equation}
where $\tilde{u}\in\mathbb{Z}_{2^q}^n$ and 
$A_0=\cos{(2\pi /2^{q})}$.
\end{Thm}

To prove the above main theorem, we need to define a computational problem called the $\varepsilon$-random linear disequations problem. This problem plays an important role for finding the approximation of the hidden shift from the discretized inputs of the oracle function.

%-------------------------------------------------------%
%       $\varepsilon$-random linear disequations        %
%-------------------------------------------------------%

\section{$\varepsilon$-random linear disequations problem}
\label{sec:eRLD}

In this section, we first define the $\varepsilon$-nearly uniform distribution and two types of the $\varepsilon$-random linear disequations problem. This problem is a generalization of the random linear disequations problem in Ref.~\cite{Ivanyos} which is the same as the problem in the case when $\varepsilon = 0$. Furthermore, we will show that the problem on $\mathbb{Z}^n_{2^q}$ can be solved in time polynomial in $n$ with a small constant error. 

Let $G$ be a finite abelian group, and let $\chi:G \rightarrow \mathbb{C} \setminus \{0\}$ be a \emph{character} of $G$, 
that is, $\chi$ is a group homomorphism from $G$ to $(\mathbb{C},\times)$.
A set of characters forms an abelian group $G*$ under pointwise multiplication. 
Note that $G*$ is isomorphic to $G$, and 
the kernel of $\chi$ is the set of group elements on which $\chi$ has value $1$.

 \begin{Def}[$\varepsilon$-nearly uniform]
A distribution over a finite set $S$ is said to be \emph{$\varepsilon$-nearly uniform} with a real tolerance parameter $c \geq 1$  over a subset $S' \subset S$ for a small enough number $0\le\varepsilon < \frac{1}{c|S'|}$ if 
\[
\begin{cases}
\mathrm{Pr}(x) \leq \varepsilon & \text{if $x \in S \setminus S'$}, \\
\frac{1}{c|S'|} \leq \mathrm{Pr}(x) \leq \frac{c}{|S'|} & 
\text{if $x \in S'$}.
\end{cases}
\]
 \end{Def}
 
Note that when $\varepsilon = 0$, an $\varepsilon$-nearly uniform distribution over $S' \in S$ is  nearly uniform over $S'$. For any $\varepsilon \ge 0$, $\varepsilon$-near uniformity over the whole set $S$ is exactly same with near uniformity over the whole set $S$.

\begin{Def}{$\varepsilon$-Random Linear Disequations$(G,c)$ - search version}
\begin{itemize}
 \item \emph{Oracle Input}: Sample from a distribution over $G*$ which is $\varepsilon$-nearly uniform with $c$ on $\{ \chi \in$ $G*$ $|$ $u$ $\notin \ker \chi \}$ for a fixed element $u$.
 \item \emph{Task}: Find the set of such elements $u$.
 \end{itemize}
\end{Def}

\begin{Def}{$\varepsilon$-Random Linear Disequations$(G,c)$ - decision version}
\begin{itemize}
\item \emph{Oracle Input}: Sample from a distribution over $G*$ which is 
\begin{itemize}
\item[-] either $\varepsilon$-nearly uniform over $\{ \chi \in$ $G*$ $|$ $u$ $\notin \ker \chi \}$ 
for a fixed element $u$
\item[-] or $\varepsilon$-nearly uniform over the whole $G$.
\end{itemize}
\item \emph{Task}: Decide which is the case.
\end{itemize}
\end{Def}

For simplicity, we denote the search version and the decision version of 
the $\varepsilon$-Random Linear Disequations$(G,c)$ 
by $\varepsilon$-$\mathrm{RLD}_s(G,c)$ and $\varepsilon$-$\mathrm{RLD}_d(G,c)$, respectively. 
 
We now take $G=\mathbb{Z}_{2^q}^n$ for integers $q$ and $n$, and 
recall that $\varepsilon$-near uniformity and near uniformity are equivalent over the whole group $\mathbb{Z}^n_{2^q}$.
If we can solve $\varepsilon$-$\mathrm{RLD}_d(H,2c)$ over subgroups $H$ of $\mathbb{Z}^n_{2^q}$, 
we decide which subgroup of $\mathbb{Z}^n_{2^q}$ contains the hidden shift $u$ by using $\varepsilon$-$\mathrm{RLD}_d(H_i,2c)$ for all maximal subgroups $H_i$ of $\mathbb{Z}^n_{2^q}$. 
Thus, we can apply the same argument in Ref.~\cite{Ivanyos} except we use $\varepsilon$-near uniformity instead of near uniformity to obtain the following proposition.

\begin{Prop}
\label{prop:reduction}
The search version of the $\varepsilon$-Random Linear Disequations over $\mathbb{Z}^n_{2^q}$ with tolerance parameter $c$,
$\varepsilon$-$\mathrm{RLD}_s(\mathbb{Z}^n_{2^q},c)$, can be reduced to $O(\mathrm{poly}(qn))$ instances of
the decision version of the $\varepsilon$-Random Linear Disequations over subgroups $H$ of $\mathbb{Z}^n_{2^q}$ with tolerance parameter $2c$,
$\varepsilon$-$\mathrm{RLD}_d(H,2c)$, in time $\mathrm{poly}(qn)$.
\end{Prop}

Now, it remains to prove that $\varepsilon$-$\mathrm{RLD}_d(\mathbb{Z}^n_{2^q},c)$ can be solved in time polynomial in $n$.
With the same observation in Ref.~\cite{Ivanyos}, 
it can be shown that deciding either a nearly uniform distribution on the whole group $\mathbb{Z}^n_{2^q}$ or on the subgroup consisting of all the elements which occur with probability at least $\varepsilon$ is equivalent to decide the existence of a polynomial with certain total degree which has the input samples as its zeros. If the distribution is the former case, we can always find such polynomial while the latter case is decidable in time polynomial in $n$ with a small constant error. 
Hence we have the following proposition which is almost the same as that in Ref.~\cite{Ivanyos}.

\begin{Prop}
\label{prop:decision}
$\varepsilon$-$\mathrm{RLD}_d(\mathbb{Z}^n_{2^q},c)$ can be solved in time $c(2nq)^{O(4^q)}$ with one-sided error $1/3$. If $q$ and $c$ are fixed, $\varepsilon$-$\mathrm{RLD}_d(\mathbb{Z}^n_{2^q},c)$ can be solved in time polynomial in $n$.
\end{Prop}

%-------------------------------------------------------------%
%  Algorithm for the continuous hidden shift problem on R^n   %
%-------------------------------------------------------------%

\section{Quantum algorithm for the continuous hidden shift problem on $\mathbb{R}^n$}
\label{sec:algorithm}

For convenience, we assume that $\Delta=\sqrt{2^q}$ and $\delta=\Delta^{-1}$ for sufficiently large $q$ with $u_i^2\le \Delta$ for all $i$ 
when $u=(u_1,u_2,\ldots, u_n)\in\mathbb{R}^n$ is the hidden shift in Definition~\ref{def:cHSP}. 
Now, we are ready to construct our algorithm for solving the continuous hidden shift problem as follows.

\begin{algorithm}[Continuous hidden shift problem on $\mathbb{R}^n$] 
$\ $ \\
\emph{Input}: The oracle function $f: \mathbb{R}^n \times \mathbb{Z}_2 \rightarrow S$ with the parameters $(\alpha, r, \xi)$ that hides the shift $u=(u_1,\dots,u_n)$ in $\mathbb{R}^n$.

\begin{enumerate}
\setcounter{enumi}{-1}
\item If $f(0,0)=f(0,1)$, then output $0$.

\item Prepare the initial state

\[
\frac{1}{\sqrt{2^{qn+1}}} 
\sum_{\tilde{x} \in \mathbb{Z}_{2^q}^n}\sum_{c \in \mathbb{Z}_2}
\ket{\tilde{x}} \ket{c}\ket{0}.
\]

\item \label{step2} Apply the unitary operation corresponding to evaluation of the oracle function $f$ to the state. 
Then the resulting state becomes

$$
\ket{\psi_\delta} = \frac{1}{\sqrt{2^{qn+1}}}
\sum_{\tilde{x} \in \mathbb{Z}_{2^q}^n} \sum_{c \in \mathbb{Z}_2}
\ket{\tilde{x}} \ket{c} \ket{f(\delta \tilde{x},c)}.
$$

\item\label{step3} Perform the quantum Fourier transform, $QFT_{{\mathbb{Z}^n_{2^q}} \times \mathbb{Z}_2}$, and measure on the first two registers.

\item Consider the samples $(\tilde{y},1)$.

\item\label{step5} Use the values of $\tilde{y}$ which are non-orthogonal to $\tilde{u}$ among the samples $(\tilde{y},1)$ 
to find $\tilde{u}$.

\item Approximate $u$ with $\delta \tilde{u}$.
\end{enumerate}

\noindent\emph{Output}:  $\delta \tilde{u} \approx u$ 

 \end{algorithm} 
 
 \begin{Rem}
We give explanations about each step in the above algorithm as follows.
  
\begin{itemize}
 
\item 

In Step~\ref{step2}, the unitary operator $U_f$ can be defined as follows. For all $\tilde{x} \in  \mathbb{Z}^n$ and $c \in \mathbb{Z}_2$, 
$$
U_f: \ket{\tilde{x}}\ket{c}\ket{k} \rightarrow 
\ket{\tilde{x}}\ket{c}\ket{f_k(\delta \tilde{x},c)}
$$
with $f_0(x,c)=f(x,c)$ and $\inn{f_k(x,c))}{f_l(x,c)}=\delta_{kl}$. We can easily see that $U_f$ is unitary.

\item
 In Step~\ref{step3}, $QFT_{\mathbb{Z}^n_{2^{q}} \times \mathbb{Z}_2}$ means the quantum Fourier transform performing over the group $\mathbb{Z}^n_{2^{q}} \times \mathbb{Z}_2$.  
After applying $QFT_{\mathbb{Z}^n_{2^{q}} \times \mathbb{Z}_2}$, the state becomes

$$
\frac{1}{2^{qn+1}} \sum _{\tilde{x},\tilde{y} \in \mathbb{Z}^n_{2^{q}}} \sum_{c,d \in \mathbb{Z}_2}
  e^{2\pi i \langle \tilde{x},\tilde{y} \rangle/2^{q}} e^{2 \pi i c \cdot d /2} 
  \ket{\tilde{y}} \ket{d}\ket{f(\delta \tilde{x},c)}.
$$
 \item
 In Step~\ref{step5}, we apply the similar method in  Ref.~\cite{Ivanyos} to solve the $\varepsilon$-random-linear-disequations problem and find $\tilde{u}$ such that $\left\Arrowvert \delta\tilde{u}-u \right\Arrowvert \leq \delta/2$.
 \end{itemize}
 \end{Rem}

%----------------------------------------%
%       Analysis of our algorithm        %
%----------------------------------------%

\section{Analysis of our algorithm}
\label{sec:analysis}

In this section, we analyze our quantum algorithm presented in the above section.  
We can precisely estimate the following $p_\delta$ which is from the Fourier transform of $\psi_\delta$: 
% that is, $\hat{\psi_\delta}=QFT_{ \mathbb{Z}^n_{2^{q}}}\psi_\delta$
\begin{eqnarray*}
 p_\delta(\tilde{y},1)&=&\inn{\hat{\psi_\delta}(\tilde{y},1)}{\hat{\psi_\delta}(\tilde{y},1)},
\end{eqnarray*}
where 
$$
\ket{\hat{\psi_\delta}(\tilde{y},1)}
= \frac{1}{2^{qn+1}} \sum _{\tilde{x} \in \mathbb{Z}^n_{2^{q}}} \sum_{c \in \mathbb{Z}_2}
  e^{2\pi i \langle \tilde{x},\tilde{y} \rangle/2^{q}} (-1)^c
  \ket{\tilde{y}} \ket{1}\ket{f(\delta \tilde{x},c)}.
$$
Since by the promises of the oracle function $f$, 
the following three equalities hold,
 \begin{eqnarray*}
\inn{f(\delta \tilde{x}',1)}{f(\delta \tilde{x},1)}
&=& \inn{f(\delta \tilde{x}'-u,0)}{f(\delta \tilde{x}-u,0)},\\
\inn{f(\delta \tilde{x}',0)}{f(\delta \tilde{x},1)}
&=& \inn{f(\delta \tilde{x}',0)}{f(\delta \tilde{x}-u,0)}, \\
\inn{f(\delta \tilde{x}',1)}{f(\delta \tilde{x},0)}
&=& \inn{f(\delta \tilde{x}'-u,0)}{f(\delta \tilde{x},0)},
\end{eqnarray*}
the probability $p_{\delta}(\tilde{y},1)$ becomes
\begin{eqnarray*}
 p_{\delta}(\tilde{y},1) 
 &=&
  \frac{1}{4^{qn+1}}\sum _{\tilde{x}, \tilde{x}' \in \mathbb{Z}^n_{2^{q}}} 
 e^{2\pi i \langle \tilde{x}-\tilde{x}',\tilde{y} \rangle/2^{q}}
   \inn{f(\delta \tilde{x}',0)}{f(\delta \tilde{x},0)}
   \\
  &&+ \frac{1}{4^{qn+1}}\sum _{\tilde{x}, \tilde{x}' \in \mathbb{Z}^n_{2^{q}}} 
 e^{2\pi i \langle \tilde{x}-\tilde{x}',\tilde{y} \rangle/2^{q}}
  \inn{f(\delta \tilde{x}'-u,0)}{f(\delta \tilde{x}-u,0)}
  \\
  &&- \frac{1}{4^{qn+1}}\sum _{\tilde{x}, \tilde{x}' \in \mathbb{Z}^n_{2^{q}}} 
 e^{2\pi i \langle \tilde{x}-\tilde{x}',\tilde{y} \rangle/2^{q}}
  \inn{f(\delta \tilde{x}',0)}{f(\delta \tilde{x}-u,0)}
  \\
  &&- \frac{1}{4^{qn+1}}\sum _{\tilde{x}, \tilde{x}' \in \mathbb{Z}^n_{2^{q}}} 
 e^{2\pi i \langle \tilde{x}-\tilde{x}',\tilde{y} \rangle/2^{q}}
  \inn{f(\delta \tilde{x}'-u,0)}{f(\delta \tilde{x},0)}.
  \end{eqnarray*}
 By using $\tilde{u} \in \mathbb{Z}^n_{2^{qn}}$ with $\left\Arrowvert \delta\tilde{u}-u \right\Arrowvert \leq \delta/2$, the probability $p_{\delta}(\tilde{y},1)$ can be rewritten as    
 \begin{eqnarray*}
 p_{\delta}(\tilde{y},1) 
 &=&
  \frac{1}{2 \cdot 4^{qn+1}}
  \left(
  2^{qn+1} 
  + \sum _{\tilde{x} \neq \tilde{x}'} 
    \cos{(2\pi \langle \tilde{x}-\tilde{x}',\tilde{y} \rangle/2^{q})}
  \inn{f(\delta \tilde{x}',0)}{f(\delta \tilde{x},0)} \right)
  \\
  && + \frac{1}{2 \cdot 4^{qn+1}}
  \sum _{\tilde{x} \neq \tilde{x}'} 
    \cos{(2\pi \langle \tilde{x}-\tilde{x}',\tilde{y} \rangle/2^{q})}
  \left( \inn{f(\delta \tilde{x}'-u,0)}{f(\delta \tilde{x}-u,0)}
  \right)
   \\
  &&- \frac{1}{2\cdot4^{qn+1}}
  \left(
  2\sum _{\tilde{x}=\tilde{u}+ \tilde{x}'} 
    \cos{(2\pi \langle \tilde{u},\tilde{y} \rangle/2^{q})}
  \inn{f(\delta \tilde{x}',0)}{f(\delta \tilde{x}'+\delta\tilde{u}-u,0)}
  \right) \\
  &&- \frac{1}{2\cdot4^{qn+1}}
  \left(
  2\sum _{\tilde{x} \neq \tilde{u}+ \tilde{x}'} 
    \cos{(2\pi \langle \tilde{x}-\tilde{x}',\tilde{y} \rangle/2^{q})}
  \inn{f(\delta \tilde{x}',0)}{f(\delta \tilde{x}-u,0)}
  \right).
   \end{eqnarray*}
%where $\tilde{u} \in \mathbb{Z}^n_{2^{qn}}$ with $\left\Arrowvert \delta\tilde{u}-u \right\Arrowvert \leq \delta/2$. 
For the convenience of calculation, let 
$
A = \cos{(2\pi \langle \tilde{u},\tilde{y} \rangle/2^{q})}.
$
It follows from the triangle inequality and the properties of the oracle function $f$ that 
\begin{eqnarray}
p_{\delta}(\tilde{y},1) 
 &\leq& \nonumber
  \frac{1}{4^{qn+1}} 
  \left(
    2^{qn}+2^{qn}(2^{qn}-1)\xi
    -(2^q-\tilde{u_i})^n A 
     \left(
     1-\frac{\alpha^2}{2^{q+3}}
     \right)
     ^{\frac{|A|+A}{2A}} \right) \\
     &&+\frac{1}{4^{qn+1}}  \cdot 2^{qn}(2^{qn}-(2^q-\tilde{u_i})^n)\xi \nonumber \\
   &\leq& \nonumber
  \frac{1}{4^{qn+1}}
  \left(
    2^{qn}
    -(2^q-2^{\frac{3q}{4}})^n A 
     \left(
     1-\frac{\alpha^2}{2^{q+3}}
     \right)
     ^{\frac{|A|+A}{2A}}
     \right)  \\
     &&+\frac{1}{4^{qn+1}}  \cdot 2^{qn}(2^{qn+1}-(2^q-2^{\frac{3q}{4}})^n)\xi
  \nonumber\\
  &\leq&
  \frac{1}{4\cdot 2^{qn}}
  \left(
    1
    -(1-2^{-\frac{q}{4}})^n A 
     \left(
     1-\frac{\alpha^2}{2^{q+3}}
     \right)
     ^{\frac{|A|+A}{2A}}
     \right)
     \nonumber \\
     &&+\frac{1}{4}  \cdot %2^{qn}
     (2-(1-2^{-\frac{q}{4}})^n)\xi. 
     \nonumber\\
 &\leq&
  \label{eq:upp}
  \frac{1}{4\cdot 2^{qn}}
  \left(
    1
    -(1-2^{-\frac{q}{4}})^n A 
     \left(
     1-\frac{\alpha^2}{2^{q+3}}
     \right)
     ^{\frac{|A|+A}{2A}}
     \right)
     +\frac{1}{2} \xi.
\end{eqnarray}
We obtain the first inequality by the condition 2 and 3 of the oracle function $f$ in Definition~\ref{def:cHSP}.
The last inequality follows from the condition of $u$ and $\tilde{u}$, that is,
$u \leq \Delta$ and $\left\Arrowvert \delta\tilde{u}-u \right\Arrowvert \leq \delta/2$. 
Note that we define the value 
$
A\left(1-\frac{\alpha^2}{2^{q+3}}
     \right)
     ^{\frac{|A|+A}{2A}}$ as zero when $A$ is zero.

Similarly, we also have
\begin{eqnarray}
p_{\delta}(\tilde{y},1) 
 &\geq& 
  \frac{1}{4^{qn+1}} 
  \left(
    2^{qn}+2^{qn}(2^{qn}-1)(-\xi)
    -(2^q-\tilde{u_i})^n A 
    \left( 1-\frac{\alpha^2}{2^{q+3}}
    \right)^{\frac{|A|+A}{2A}}
   \right) \nonumber\\
   &&-\frac{1}{4^{qn+1}} 
  \left(
     2^{qn}(2^{qn}-(2^q-\tilde{u_i})^n)\xi
  \right) \nonumber\\
  &\geq& 
  \frac{1}{4^{qn+1}} 
  \left(
    2^{qn}
    - A\left(
    \left(2^q-2^{\frac{3q}{4}}\right)^n
    \left( 1-\frac{\alpha^2}{2^{q+3}}
    \right)
    \right)^{\frac{|A|+A}{2A}}
    \right)\nonumber\\
     &&-\frac{1}{4^{qn+1}} \left(2^{qn}(2^{qn}-1+2^{qn}-(2^q-0)^n)\xi
  \right) \nonumber\\
  &\geq& 
  \frac{1}{4\cdot 2^{qn}}
  \left(
    1-A \left(
    \left(1-2^{-\frac{q}{4}}\right)^n
    \left( 1-\frac{\alpha^2}{2^{q+3}}
    \right)
    \right)^{\frac{|A|+A}{2A}}
    -(2^{qn}-1)\xi
  \right)
     \nonumber \\
   &\geq& 
  \frac{1}{4\cdot 2^{qn}}
  \left(
    1-A \left(
    \left(1-2^{-\frac{q}{4}}\right)^n
    \left( 1-\frac{\alpha^2}{2^{q+3}}
    \right)
    \right)^{\frac{|A|+A}{2A}}
    \right)
    -\frac{1}{4 
    }\xi.
  \label{eq:low}
\end{eqnarray}
Here, we also define the value
    $A \left(
    \left(1-2^{-\frac{q}{4}}\right)^n
    \left( 1-\frac{\alpha^2}{2^{q+3}}
    \right)
    \right)^{\frac{|A|+A}{2A}}$ 
     as zero when $A$ is zero.

Note that the probability that a sample $\tilde{y}$ is orthogonal to $\tilde{u}$ is not zero,  
 which is different from the original hidden shift problem. 
 However, it can be shown that the probability is small 
 by exploiting the Lipschitz condition of the oracle function $f$. 
 In particular, we can also prove that the samples $(\tilde{y},1)$ after the Fourier sampling subroutine are mostly non-orthogonal to $\tilde{u}$ when $q$ is a multiple of 4 (See Appendix~\ref{sec:maxpf} for the details).
 
 In our algorithm, the samples $\tilde{y}$ are elements in the finite abelian group $\mathbb{Z}^n_{2^q}$, and the probability that a sample $\tilde{y}$ is orthogonal to $\tilde{u}$ is not zero even though it is negligible. This causes an obstacle to apply the method used in Ref.~\cite{Ivanyos} directly to our algorithm. In order to handle this difficulty, we have introduced the concept of the \emph{$\varepsilon$-nearly uniform distribution} in Sec.~\ref{sec:eRLD}, 
 and we now show that the sample distribution is  $\varepsilon$-nearly uniform with a certain tolerance $c$ if the parameters $(\alpha, r, \xi)$ of the oracle function $f$ satisfy the following conditions.
 
 Let $q$ be an integer such that $\alpha^2 < 2^{q+3}$. Then it can be shown that the values in Eq.~(\ref{eq:upp}) and Eq.~(\ref{eq:low}) are positive. Suppose that the oracle function has the parameters $(\alpha, r, \xi)$ such that 
 $r \le \frac{\delta}{2}=\frac{1}{\sqrt{2^{q+2}}}$ and $\xi$ be a non-negative number less than both of the values in Eq.~(\ref{xi_1}) and Eq.~(\ref{xi_2}). 
 Then we obtain the following theorem.

\begin{Thm}
\label{thm:e-linearly_unif}
%%% condition of \xi %%% 
Let $q$ be an integer 
such that $\alpha^2 < 2^{q+3}$,
where $\alpha$ is the Lipschitz constant of the oracle function $f$ with the parameters $(\alpha, r, \xi)$, and let $\delta = \frac{1}{\sqrt{2^q}}$. Let $S'$ be the subset of $\mathbb{Z}^n_{2^q}$ consisting of all the elements which are not orthogonal to the specific element $\tilde{u} \in \mathbb{Z}^n_{2^q}$ with $\left\Arrowvert \delta\tilde{u}-u \right\Arrowvert \leq \delta/2$. 
Assume that $r= \frac{\delta}{2}$ and $\xi$ is a positive number less than both of the values in Eq.~(\ref{xi_1}) and Eq.~(\ref{xi_2}).
Then there exists a real number $c > 1$ such that the distribution of the samples $\tilde{y}$ in our algorithm is $\varepsilon$-nearly uniform over $S'$ with tolerance $c$, 
where 
\[
\varepsilon = \frac{1
    -\left(1-2^{-\frac{q}{4}}\right)^n
     \left(
     1-\frac{\alpha^2}{2^{q+3}}
     \right)
     +2^{qn+1} \xi}
     {4 \cdot 2^{qn}}.
\]
\end{Thm}

\proof{
Let 
$
A=\cos({2 \pi \langle \tilde{u}, \tilde{y} \rangle /2^q}).
$
%
% We now let $\tilde{y} \in S'$, that is,  $\left<\tilde{u},\tilde{y}\right> \neq 0$.
% Then $A \neq 1$.
Then $-1 \le A \le 1$.
If $\langle \tilde{u}, \tilde{y} \rangle \neq 0$, by the assumption of $\xi$, we can show that 
\begin{equation}
    \frac{4 \cdot 2^{qn}} 
    {|S'|\left( 1-A\left(
    \left(1-2^{-\frac{q}{4}}\right)^n
    \left( 1-\frac{\alpha^2}{2^{q+3}}
    \right)
    \right)^{\frac{|A|+A}{A}}-2^{qn}\xi
    \right)}
    \label{eq:bound_c_low}
\end{equation}
 is less than 
 \begin{equation}
 \frac{4 \cdot 2^{qn}}
    {|S'|\left(
    1- \left(1-2^{-\frac{q}{4}}\right)^n
    \left( 1-\frac{\alpha^2}{2^{q+3}}
    \right)
    + 2^{qn+1}\xi
 \right)}. 
 \label{eq:bound_c_upp}
\end{equation}
Indeed, since $\xi$ is less than or equal to the value in Eq.~(\ref{xi_2}) which is non-zero when $\langle \tilde{u}, \tilde{y} \rangle \neq 0$, we have
\begin{equation}
    0< \xi < 
    \frac{
    \left(1-2^{-\frac{q}{4}}\right)^n
    \left( 1-\frac{\alpha^2}{2^{q+3}}
    \right)
    - A
    \left(
    \left(1-2^{-\frac{q}{4}}\right)^n
    \left( 1-\frac{\alpha^2}{2^{q+3}}
    \right)
    \right)^{\frac{|A|+A}{A}}}
    {3 \cdot 2^{qn}},
\end{equation}
or, equivalently, 
\begin{eqnarray}
    3 \cdot 2^{qn} \xi
     < 
     \left(1-2^{-\frac{q}{4}}\right)^n
    \left( 1-\frac{\alpha^2}{2^{q+3}}
    \right)
    - A
    \left(
    \left(1-2^{-\frac{q}{4}}\right)^n
    \left( 1-\frac{\alpha^2}{2^{q+3}}
    \right)
    \right)^{\frac{|A|+A}{A}}.
\end{eqnarray}
This inequality is also equivalent to 
\begin{eqnarray}
     1&-&\left(1-2^{-\frac{q}{4}}\right)^n
    \left( 1-\frac{\alpha^2}{2^{q+3}}
    \right)
    + 2^{qn+1}\xi \nonumber
    \\
    &<& 1-A\left(
    \left(1-2^{-\frac{q}{4}}\right)^n
    \left( 1-\frac{\alpha^2}{2^{q+3}}
    \right)
    \right)^{\frac{|A|+A}{A}}-2^{qn}\xi,
\end{eqnarray}
which directly leads to the inequality we want to show.
Hence, we can choose any real number $c$ between those two values in Eq.~(\ref{eq:bound_c_low}) and Eq.~(\ref{eq:bound_c_upp}).

Note that $c$ is greater than $1$ because
\begin{eqnarray*}
c &>& \frac{4 \cdot 2^{qn}}
    {
    |S'|\left( 1-A\left(
    \left(1-2^{-\frac{q}{4}}\right)^n
    \left( 1-\frac{\alpha^2}{2^{q+3}}
    \right)
    \right)^{\frac{|A|+A}{2A}}-2^{qn}\xi
    \right)
    }
     \\
  &>&
  \frac{4 \cdot 2^{qn}}
    {
    2^{qn} \left( 1-A\left(
    \left(1-2^{-\frac{q}{4}}\right)^n
    \left( 1-\frac{\alpha^2}{2^{q+3}}
    \right)
    \right)^{\frac{|A|+A}{2A}}-2^{qn}\xi
    \right)
    } \\
  &>& \frac{4}{1+1-0}=2>1.
\end{eqnarray*}
Now, we want to show that the distribution of the samples $\tilde{y}$ in our algorithm is $\varepsilon$-nearly uniform over $S'$ with such $c$ for some $\varepsilon$.

Since $c$ is greater than the value in Eq.~(\ref{eq:bound_c_low}), it follows from the inequality in Eq.~(\ref{eq:upp}) that
\begin{eqnarray}
\label{eq:lower}
\frac{1}{c|S'|} &<& 
\frac{1}{4\cdot 2^{qn}}
  \left(
    1-A \left(
    \left(1-2^{-\frac{q}{4}}\right)^n
    \left( 1-\frac{\alpha^2}{2^{q+3}}
    \right)
    \right)^{\frac{|A|+A}{2A}}
    -2^{qn}\xi
  \right) \nonumber\\
  &\le& p_\delta(\tilde{y},1).
\end{eqnarray}
On the other hand, since $\xi$ is less than the value in Eq.~(\ref{xi_1}), by combining the inequality in Eq.~(\ref{eq:low}), we also obtain that 
\begin{eqnarray}
p_\delta(\tilde{y},1) &\le&
\frac{1}{4\cdot 2^{qn}}
  \left(
    1
    -(1-2^{-\frac{q}{4}})^n A 
     \left(
     1-\frac{\alpha^2}{2^{q+3}}
     \right)
     ^{\frac{|A|+A}{2A}}
     +2^{qn+1}\xi
  \right) \nonumber \\
  &<& \frac{1}{2^{qn}} < \frac{c}{|S'|}
  \label{eq:upper},
\end{eqnarray}
where the second inequality comes from Eq.~(\ref{xi_1}) and the inequality
\[
A \left(
     1-\frac{\alpha^2}{2^{q+3}}
   \right)
    ^{\frac{|A|+A}{2A}} + 1 >0.
\]
Thus, 
$$
\frac{1}{c|S'|} \le p_\delta(\tilde{y},1) \le \frac{c}{|S'|}
$$
for $\tilde{y} \in S'$.
Now, let us consider
$$
\varepsilon = \frac{1
    -\left(1-2^{-\frac{q}{4}}\right)^n
     \left(
     1-\frac{\alpha^2}{2^{q+3}}
     \right)
     +2^{qn+1}\xi}
     {4 \cdot 2^{qn}}.
$$
Then $\varepsilon$ is clearly positive. Note that $A=\cos({2 \pi \langle \tilde{u}, \tilde{y} \rangle /2^q})=1$ for $\tilde{y}\in S-S'$, that is, $\left<\tilde{u},\tilde{y}\right>=0$.
 So, the value in Eq.~(\ref{xi_2}) must be zero which implies that $\xi=0$.
 Hence, in this case, the inequalities in Eq.~(\ref{eq:upp}) and Eq.~(\ref{eq:low}) are saturated. 
 $$
 p_\delta(\tilde{y},1) =
\frac{1}{4\cdot 2^{qn}}
  \left(
    1
    -(1-2^{-\frac{q}{4}})^n 
     \left(
     1-\frac{\alpha^2}{2^{q+3}}
     \right)
     \right)
     \leq \varepsilon.
 $$
Moreover, from the second inequality in Eq.~(\ref{eq:bound_c_upp}), we also have  
\begin{equation}
    \varepsilon < \frac{1}{c|S'|} 
    \label{eq:epsilon_upper}
\end{equation}
since 
\begin{equation}
c <
 \frac{4 \cdot 2^{qn}}
    {|S'|\left(
    1- \left(1-2^{-\frac{q}{4}}\right)^n
    \left( 1-\frac{\alpha^2}{2^{q+3}}
    \right)
    + 2^{qn+1}\xi
 \right)} = \frac{1}{\varepsilon |S'|}.
\nonumber
\end{equation}

Therefore, it follows from the inequalities in Eqs.~(\ref{eq:lower}), (\ref{eq:upper}) and  (\ref{eq:epsilon_upper}) that the distribution of the samples $\tilde{y}$ in our algorithm is $\varepsilon$-nearly uniform over $S'$ with tolerance $c$.
\hfill$\square$

We can now prove the main theorem as follows. 

\vspace{2mm}
\noindent\textit{Proof of Theorem~\ref{thm:main}}
We can perform $QFT_{\mathbb{Z}^n_{2^{q}} \times \mathbb{Z}_2}$ in time $O(n^2)$ and obtain the Fourier samples $\tilde{y}$. 
By Theorem~\ref{thm:e-linearly_unif}, the distribution of the samples $\tilde{y}$ in our algorithm is $\varepsilon$-nearly uniform with a certain tolerance $c$ over the subset consisting of all the samples that are non-orthogonal to $\tilde{u}$. Thus, we obtain an instance of $\varepsilon$-$\mathrm{RLD}_s(\mathbb{Z}^n_{2^q},c)$ from the samples in the algorithm and we get $\tilde{u}$ as the solution of  $\varepsilon$-$\mathrm{RLD}_s(\mathbb{Z}^n_{2^q},c)$.
Since $q$ and $c$ are fixed in our algorithm, it follows from  Proposition~\ref{prop:reduction} and Proposition~\ref{prop:decision} that 
$\varepsilon$-$\mathrm{RLD}_s(\mathbb{Z}^n_{2^q},c)$ is reducible to 
$\varepsilon$-$\mathrm{RLD}_d(\mathbb{Z}^n_{2^q},2c)$ which can be solved in time polynomial in $n$.
Thus, we can find the approximation $\delta \tilde{u}$ of the hidden shift $u$ in time polynomial in $n$ by solving $\varepsilon$-$\mathrm{RLD}_d(\mathbb{Z}^n_{2^q},2c)$ for the tolerance $c$.
\hfill$\square$
}

\begin{Rem}
If $u$ lies on the $\delta$-grid, that is, $u=\delta \tilde{u}$, then  
$1-\frac{\alpha^2}{2^{q+3}}$
in the above all equations and all inequalities 
can be replaced by $1$.
So, we can more briefly prove that the distribution of the samples $\tilde{y}$ in our algorithm is $\varepsilon$-nearly uniform over $S'$ with tolerance $c$. 
Then we have an instance of the search version of $\varepsilon$-$\mathrm{RLD}(\mathbb{Z}^n_{2^q},c)$ and $\tilde{u}$ can be obtained by solving this problem. 
It follows from Proposition~\ref{prop:reduction} that we can reduce the search version of $\varepsilon$-$\mathrm{RLD}(\mathbb{Z}^n_{2^q},c)$ to the decision version of $\varepsilon$-$\mathrm{RLD}(\mathbb{Z}^n_{2^q},2c)$ in time polynomial in $n$. 
Moreover, since $q$ and $c$ are fixed, by Proposition~\ref{prop:decision}, 
$\varepsilon$-$\mathrm{RLD}_d(\mathbb{Z}^n_{2^q},2c)$ can be solved in time polynomial in $n$. 
Thus, we can efficiently find the $\frac{\delta}{2}$-approximation $\tilde{u}$ of $u$ by solving $\varepsilon$-$\mathrm{RLD}_d(\mathbb{Z}^n_{2^q},2c)$.
Obviously, Theorem~\ref{thm:main} and Theorem \ref{thm:e-linearly_unif} can be considered as general versions of this case.
\end{Rem}

%-----------------------%
%       Conclusion      %
%-----------------------%

\section{Conclusion and discussion}
\label{sec:discussion}

 In this work, we have presented the continuous hidden shift problem over the continuous group $\mathbb{R}^n$, and 
have defined two types of the $\varepsilon$-random linear disequations problem, $\varepsilon$-$\mathrm{RLD}_s(G,c)$ and $\varepsilon$-$\mathrm{RLD}_d(G,c)$, in order to employ the similar approach in Ref.~\cite{Ivanyos}
for finding the hidden shift in $\mathbb{Z}_{p^k}^n$.
We have also shown that a quantum computer can efficiently solve the problem in time polynomial in $n$ by solving $\varepsilon$-$\mathrm{RLD}_d(\mathbb{Z}^n_{2^q},2c)$.

 It has been known that the hidden shift problem over discrete groups can affect cryptosystems~\cite{AR17,BN18}.
 In particular, Bonnetain and Naya-Plasencia~\cite{BN18} recently constructed an efficient quantum algorithm 
 to solve the hidden shift problem over the abelian group $\mathbb{Z}_{2^p}^w$, and 
 proved that the algorithm can be used to establish a quantum attack in a cryptosystem
 claimed to be secure quantumly. 
  Thus, it is natural to consider the question about whether the continuous version of the hidden shift problem can also have any crypto-related applications. In fact, a similar consideration has been involved in the previous results.
 In Refs.~\cite{EHKS14,BS16}, it has been shown that an efficient quantum algorithm for the continuous hidden subgroup problem on $\mathbb{R}^n$ induces a quantum attack on cryptosystems based on the hardness of finding a short generator of a principal ideal, although the original hidden subgroup problem on $\mathbb{R}^n$ cannot provide such an attack to break them~\cite{Hal05,Hal07}. 
 Inspired by these results, we expect that our result could have applications related to cryptography.

 Furthermore, we can also try to consider continuous versions of other algebraic problems with hidden structure such as hidden symmetry subgroup problem, hidden polynomial problem, and so on.

%----------------------------------------%
%            Acknowledgements            %
%----------------------------------------%

\acknowledgements{
We would like to thank Fang Song for fruitful discussion.
This research was supported by the National Research Foundation of Korea grant funded by the Ministry of Science and ICT (MSIT) (Grant no. NRF-2019R1A2C1006337) and (Grant no. NRF-2020M3E4A1079678). E.B. acknowledges support from  the National Research Foundation of Korea grant funded by the MSIT (Grant no. NRF-2019K1A3A1A12071493), 
and 
S.L. acknowledges support from the MSIT, under the Information Technology Research Center support program (IITP-2021-2018-0-01402) supervised by the Institute for Information \& Communications Technology Planning \& Evaluation, and the Quantum Information Science and Technologies program of the National Research Foundation of Korea funded by the MSIT (Grant no. NRF-2020M3H3A1105796).
}

\bibliography{ctsHTP}

%----------------------------------------%
%                Appendix                %
%----------------------------------------%
\appendix

\section{The sample $\tilde{y}$ is mostly orthogonal to $\tilde{u}$}
\label{sec:maxpf}
In this section, we show that the probability that a sample $\tilde{y}$ is orthogonal to $\tilde{u}$ is small enough when $q$ is a sufficiently large multiple of 4.
In order to do that, 
we consider the case when the number of $\tilde{y}$'s satisfying the equation $\left< \tilde{u},\tilde{y}\right>=0$ 
attains a maximum value. 
Note that we get an approximation value $\delta\tilde{u}$ of $u$ with $\left\Arrowvert \delta\tilde{u}-u \right\Arrowvert \leq \delta/2$ 
by means of our algorithm in Sec.~\ref{sec:algorithm}. 

\begin{Prop}
\label{lem:maxprob}
Let $q=4k$ for a positive integer $k$ and let $\Delta=\sqrt{2^q}$ such that $u_i^2 \leq \Delta$ for all $i$.
Then $\tilde{u}=(\tilde{u}_1,\tilde{u}_2,\ldots, \tilde{u}_n) \in \mathbb{Z}_{2^q}^n$ satisfies $\tilde{u}_i \leq 2^{3k}$ for all $i$.
The number of $\tilde{y}$'s which are orthogonal to $\tilde{u}$ in $\mathbb{Z}_{2^q}^n$ 
has the maximum value, $2^{k(4n-1)}$,  
when $\tilde{u}=(2^{3k},2^{3k},\dots,2^{3k})$.
Moreover, the probability that $\tilde{y} \in \mathbb{Z}_{2^q}^n$ is orthogonal to 
$\tilde{u}$ is at most $1/2^{k}$ in our algorithm.
\end{Prop}

In order to prove Proposition~\ref{lem:maxprob}, 
we first show that if $\tilde{u}$ has the same coordinates, 
then the number of $\tilde{y}$'s satisfying the equation $\left< \tilde{u},\tilde{y}\right>=0$ 
in $\mathbb{Z}_{2^q}^n$ becomes less than or equal to
the number in the case that we make one of the same coordinates of $\tilde{u}$ twice.
 
\begin{Lem}
\label{prop:2times}
For any positive integers $n$ and $k$, let $q=4k$. 
The number of $\tilde{y}$'s which are orthogonal to $\tilde{u}=(\tilde{u}_1,\dots,\tilde{u}_n)$ with $\tilde{u}_i =\tilde{u}_j$ 
for some $i \neq j$,  
is less than or equal to the number of $\tilde{y}$'s  
which are orthogonal to $\tilde{u}'=(\tilde{u}_1,\dots,2\tilde{u}_i,\dots,\tilde{u}_n)$
or $(\tilde{u}_1,\dots,2\tilde{u}_j,\dots,\tilde{u}_n)$.

\end{Lem}

\proof{
Without loss of generality, we may assume that $\tilde{u}_1=\tilde{u}_2$. 
We want to show that the number of $\tilde{y}$ satisfying $\left< \tilde{u},\tilde{y} \right>\equiv 0 \pmod  {2^q}$ 
for $\tilde{u}=(\tilde{u}_1,\tilde{u}_1,\tilde{u}_3,\dots,\tilde{u}_n)$ is less than or equal to the number of $\tilde{y}$ satisfying $\left< \tilde{u}',\tilde{y} \right>\equiv 0 \pmod  {2^q}$ for $\tilde{u}'=(2\tilde{u}_1,\tilde{u}_1,\tilde{u}_3,\dots,\tilde{u}_n)$.
We can establish an injective function from the solutions 
$\tilde{y}=(\tilde{y}_1,\dots,\tilde{y}_n)$ of the equation $\left< \tilde{u},\tilde{y} \right>=\sum_{i=1}^n\tilde{u}_i\tilde{y}_i\equiv 0 \pmod  {2^q}$ to the solutions $\tilde{y}'=(\tilde{y}_1',\dots,\tilde{y}_n')$ of the equation $\left< \tilde{u}',\tilde{y}' \right>\equiv 0 \pmod  {2^q}$ as follows.
$$
(\tilde{y}_1,\dots,\tilde{y}_n)  \longmapsto  (\tilde{y}_1',\dots,\tilde{y}_n')=
\begin{cases}
(m_1,\tilde{y}_2,\dots,\tilde{y}_n) &  \textrm{if}~~\tilde{y}_1=2m_1, \\ 
(m_1,\tilde{y}_2-1,\dots,\tilde{y}_n) & \textrm{if}~~\tilde{y}_1=2m_1-1.
\end{cases}
$$
Then it can be easily shown that $\left< \tilde{u},\tilde{y}\right>=\left< \tilde{u}',\tilde{y}' \right>$.
In fact, if $\tilde{y}_1=2m_1$, then 
\begin{eqnarray*}
\left< \tilde{u}',\tilde{y}'\right>= 
\tilde{u}_1'm_1 + \tilde{u}_2'\tilde{y}_2'+ \cdots + \tilde{u}_n'\tilde{y}_n'=
2\tilde{u}_1 m_1 + \tilde{u}_2\tilde{y}_2+ \cdots + \tilde{u}_n\tilde{y}_n=
\left< \tilde{u},\tilde{y}\right>.
\end{eqnarray*}
Similary,  if $\tilde{y}_1=2m_1-1$, then 
\begin{eqnarray*}
\left< \tilde{u}',\tilde{y}'\right>&=& 
\tilde{u}_1'm_1 + \tilde{u}_2'(\tilde{y}_2-1)+ \cdots + \tilde{u}_n'\tilde{y}_n' \\
&=&
2\tilde{u}_1 m_1 + \tilde{u}_1\tilde{y}_2 -\tilde{u}_1 + \cdots + \tilde{u}_n\tilde{y}_n \\
&=&
\tilde{u}_1 (2m_1-1) + \tilde{u}_1\tilde{y}_2 + \cdots + \tilde{u}_n\tilde{y}_n
=\left< \tilde{u},\tilde{y}\right>.
\end{eqnarray*}
\hfill$\square$
}

\vspace{2mm}
For the next step, we show that if all coordinates of $\tilde{u}$ have the form of $2^t$, 
the number of $\tilde{y}$'s which are orthogonal to $\tilde{u}$ 
becomes less than or equal to the number in the case that we change the smallest coordinates of $\tilde{u}$ to the second smallest one
as in the following lemma.
\begin{Lem}
\label{prop:distinct}
For any positive integers $n$ and $k$, let $q=4k$ and $\tilde{u}=(2^{t_1},\dots,2^{t_n})$ 
such that $t_i \le 3k$ for all $i$ and $t_i$'s are all distinct, say $t_{i_1}< \cdots <t_{i_n}$ with $i_j \in [n]:=\{1, 2, \ldots, n\}$. 
The number of $\tilde{y}$'s 
which are orthogonal to $\tilde{u}=(2^{t_1},\dots,2^{t_n})$ in $\mathbb{Z}_{2^q}^n$  
is less than or equal to the number of $\tilde{y}$'s  
which are orthogonal to $\tilde{u}'=(\tilde{u}_1',\dots,\tilde{u}_n')$ in $\mathbb{Z}_{2^q}^n$ with 
$$
\tilde{u}_{i}'=
\begin{cases}
2^{t_{i_2}} & i =i_1 , \\ 2^{t_i}  & i \neq i_1.
\end{cases}
$$

\end{Lem}

\proof{
Suppose that $\tilde{u}_i'$'s are  all distinct. 
Without loss of generality, assume that $t_1 < t_2 < \cdots < t_n$, where $t_i \le 3k$ for all $i$.
Then we can construct an injective functions from the solutions 
$\tilde{y}=(\tilde{y}_1,\dots,\tilde{y}_n)$ of the equation $\left< \tilde{u},\tilde{y} \right>\equiv 0 \pmod  {2^q}$ 
for $\tilde{u}=(2^{t_1},\dots,2^{t_n})$ 
to the solutions $\tilde{y}'=(\tilde{y}_1',\dots,\tilde{y}_n')$ of the equation $\left< \tilde{u}',\tilde{y}' \right>\equiv 0 \pmod  {2^q}$ for $\tilde{u}'=(2^{t_2},2^{t_2},\dots,2^{t_n})$ 
as follows.
$$
(\tilde{y}_1,\dots,\tilde{y}_n)  \longmapsto  (\tilde{y}_1',\dots,\tilde{y}_n')=(2^{t_1-t_2}\tilde{y}_1,\tilde{y}_2,\dots,\tilde{y}_n).
$$
Note that $\tilde{y}_1$ is a multiple of $2^{t_2-t_1}$.
Indeed, since the condition $\left< \tilde{u},\tilde{y} \right>=0 \pmod  {2^q}$ implies that  
\begin{eqnarray*}
\tilde{y}_1+ 2^{t_2-t_1} \left( \tilde{y}_2 + 2^{t_3-t_2+t_1}\tilde{y}_3 + \dots + 2^{t_n-t_2+t_1}\tilde{y}_n \right) 
=l2^{q-t_1}
\end{eqnarray*}
for $l=0,1,\ldots,\sum_{i=1}^n \tilde{u}_i -1$, 
it is clear that $\tilde{y}_1=2^{t_2-t_1}\mu$ for some positive integer $\mu$. 
Hence, we can see that $\tilde{y}_1'=2^{t_1-t_2}\tilde{y}_1$ must be a positive integer.

In addition, we have 
\[ 
\left< \tilde{u}',\tilde{y}' \right>=2^{t_2}(2^{t_1-t_2}\tilde{y}_1)+2^{t_2}\tilde{y}_2 + \cdots +2^{t_n}\tilde{y}_n
=\left< \tilde{u},\tilde{y} \right>.
\]
\hfill$\square$
}

\vspace{2mm}
On the other hand, we can also consider the case that 
$\tilde{u}$ has at least one coordinate which cannot be written as the form of $2^t$.
In this case, we have the following lemma. 
\begin{Lem}
\label{prop:u_j}
For any positive integers $n$ and $k$, let $q=4k$ and $\tilde{u} \in \mathbb{Z}_{2^q}^n$.
If there is $j\in [n]$ such that $\tilde{u}_j=v 2^t$ with $\mathrm{gcd}(v,2)=1$ and $t \ge 0$, 
the number of $\tilde{y}$ orthogonal to $\tilde{u}$ in $\mathbb{Z}_{2^q}^n$  
is exactly the same as the number of $\tilde{y}$'s  
which are orthogonal to $\tilde{u}'=(\tilde{u}_1',\dots,\tilde{u}_n')$ in $\mathbb{Z}_{2^q}^n$ with 
$$
\tilde{u}_i'=
\begin{cases}
2^t & i=j, \\ 
\tilde{u}_i  & i \neq j.
\end{cases}
$$
\end{Lem}

\proof{
Without loss of generality, we may assume that $\tilde{u}_1=v2^t$ for some $v$ coprime with 2 and non-negative integer $t$.
Then we can construct a one-to-one correspondence between the solutions 
$\tilde{y}=(\tilde{y}_1,\dots,\tilde{y}_n)$ of the equation $\left< \tilde{u},\tilde{y} \right>\equiv 0 \pmod  {2^q}$ 
for $\tilde{u}=(v2^t,\tilde{u}_2,\dots,\tilde{u}_n)$ 
to the solutions $\tilde{y}'=(\tilde{y}_1',\dots,\tilde{y}_n')$ of the equation $\left< \tilde{u}',\tilde{y}' \right>\equiv 0 \pmod  {2^q}$ for $\tilde{u}'=(2^t,\tilde{u}_2,\dots,\tilde{u}_n)$ 
as follows.
\[
(\tilde{y}_1,\dots,\tilde{y}_n)  \longmapsto  (\tilde{y}_1',\dots,\tilde{y}_n')=
(v\tilde{y}_1,\tilde{y}_2,\dots,\tilde{y}_n), 
\]

Since $v$ is invertible in $\mathbb{Z}_{2^q}^n$, the above map is a bijection.
In addition, we clearly have
\[
\sum_{i=1}^n\tilde{u}_i\tilde{y}_i=\sum_{i=1}^n\tilde{u}'_i\tilde{y}'_i.  
\]
This completes the proof.
\hfill$\square$
}

\vspace{2mm}
Combining the above lemmas, we finally prove that 
the number of $\tilde{y}$'s orthogonal to $\tilde{u}$ attains a maximum value 
when all coordinates of $\tilde{u}$ are $2^{3k}$, which is the maximum of its each coordinate.
Thus we can get an upper bound on 
the probability that $\tilde{y}$ is orthogonal to $\tilde{u}$.

\subsection{Proof of Proposition~\ref{lem:maxprob}}
We first note that 
for any $\tilde{u}=(\tilde{u}_1,\tilde{u}_2,\ldots,\tilde{u}_n)\in\mathbb{Z}_{2^q}^n$, 
each coordinate $\tilde{u}_j$ can be expressed as $v_j2^{t_j}$ for some $v_j$ comprime to 2 and non-negative integer $t_j$.
Thus by repeatedly using Lemma~\ref{prop:u_j}, 
we can know that the number of solutions $\tilde{y}$ of the equation $\left< \tilde{u},\tilde{y} \right>\equiv 0 \pmod  {2^q}$ 
is the same as the number in the case that
$\tilde{u}=(2^{t_1},2^{t_2},\ldots,2^{t_n})$.
Furthermore, $u_i\le 2^k$ for all $i\in [n]$ by the assumption, 
and 
\[
\left|\frac{\tilde{u}_i}{2^{2k}}-u_i\right|\le \frac{1}{2^{2k+1}},
\]
since $\delta=1/2^{2k}$ and $\delta\tilde{u}$ is a $\frac{\delta}{2}$-approximation of $u$.
Thus it is clear that $\tilde{u}_i\le 2^{3k}$ for all $i\in [n]$.

Now, let us consider the case that $\tilde{u}=(2^{t_1},2^{t_2},\ldots,2^{t_n})$ with $t_i \le 3k$ for all $i$. 
Without loss of generality, we may assume that $t_1 \leq \cdots \leq t_n$.
If $t_i$'s are all distinct, it follows from Lemma~\ref{prop:distinct} that 
the number of the solutions $\tilde{y}$ of the equation $\left< \tilde{u},\tilde{y} \right>\equiv 0 \pmod  {2^q}$ is 
less than or equal to the number of the solutions in the case that $\tilde{u}=(2^{t_2},2^{t_2},2^{t_3}\ldots,2^{t_n})$. 
Therefore, by exploiting Lemma~\ref{prop:2times} and Lemma~\ref{prop:distinct} repeatedly, 
it can be shown that the number of $\tilde{y}$'s
which are orthogonal to $\tilde{u}$ in $\mathbb{Z}_{2^q}^n$ 
has the maximum value when $\tilde{u}=(2^{3k},\dots,2^{3k})$.

For the next step, we want to calculate the exact number of $\tilde{y}$'s satisfying 
$\left< \tilde{u},\tilde{y} \right>\equiv 0\pmod  {2^q}$ when $\tilde{u}=(2^{3k},\dots,2^{3k})$.
As a matter of fact, we can prove that the number of integer solutions $\tilde{y}$ 
of the equation $2^{3k}\tilde{y}_1+\cdots+ 2^{3k}\tilde{y}_n\equiv 0 \pmod  {2^q}$ is equal to 
\begin{equation}
\label{eq:Lem}
\sum_{i=0}^{2^{3k}-1}\sum_{j=0}^{n-1}
(-1)^j\binom{n-1}{j}\binom{n-1+(n-j)2^{4k}+i2^{k}}{n-1}=2^{3k}2^{4k(n-1)}=2^{k(4n-1)}
\end{equation}
for any $i$
by induction on $n \ge 1$. To do this, we need to show the following two claims.

%%Claim1%%
\begin{Claim}\label{claim1}
Let $n \ge 0$, $L$, and $l$ any fixed positive integers. 
Then 
\begin{equation}
\label{eq:downstair}
\sum_{i=0}^{n} (-1)^i\binom{n}{i}\binom{L-il}{n} = 
\sum_{i=0}^{n} (-1)^i\binom{n}{i}\binom{L-il-1}{n}.
\end{equation}
\end{Claim}

\vspace{2mm}
\noindent
\textit{Proof of Claim~\ref{claim1}}
We use the induction on $n$.
If $n=0$, the statement is obviously true. 
Now we suppose that Eq.~(\ref{eq:downstair}) holds for a certain $n\ge 0$.
From the Pascal's relation, we observe that
\begin{eqnarray*}
\sum_{i=0}^{n+1} (-1)^i\binom{n+1}{i}\binom{L-il}{n+1}
&=& \sum_{i=0}^{n+1} (-1)^i\binom{n+1}{i} \left[\binom{L-il-1}{n+1}+\binom{L-il-1}{n} \right] \\
&=& \sum_{i=0}^{n+1} (-1)^i\binom{n+1}{i}\binom{L-il-1}{n+1},
\end{eqnarray*} 
and hence Eq.~(\ref{eq:downstair}) holds for $n+1$ as well. Here, the last equality holds since 
\begin{eqnarray*}
\sum_{i=0}^{n+1} (-1)^i\binom{n+1}{i}\binom{L-il-1}{n} 
&=& \binom{n}{0}\binom{L-1}{n} 
+ (-1)^{n+1}\binom{n}{n}\binom{L-(n+1)l-1}{n} \\
&& + \sum_{i=1}^{n} (-1)^i\ \left[ \binom{n}{i}+\binom{n}{i-1} \right] \binom{L-il-1}{n} \\
&=& \sum_{i=0}^{n} (-1)^i\binom{n}{i}\binom{L-il-1}{n} - 
\sum_{i=0}^{n} (-1)^i\binom{n}{i}\binom{L-(i+1)l-1}{n} \\
&=& 0 
\end{eqnarray*} 
by employing the induction hypothesis $l$ times.
\hfill$\square$

\vspace{2mm}
This claim implies that 
for any fixed positive integers, $n \ge 0$, $L$, $L'$, and $l$,  
\begin{equation}
\label{eq:claim11}
\sum_{i=0}^{n} (-1)^i\binom{n}{i}\binom{L-il}{n} = 
\sum_{i=0}^{n} (-1)^i\binom{n}{i}\binom{L'-il}{n}.
\end{equation}
 By Claim~\ref{claim1}, we can prove the following claim, 
 which can be directly used to prove the Eq.~(\ref{eq:Lem}).

%%Claim2 %%
\begin{Claim}\label{claim2}
For each $i\in\{0,1,\ldots, 2^{3k}-1\}$ and $n \ge 1$, 
\begin{equation}
\sum_{j=0}^{n-1}
(-1)^j\binom{n-1}{j}\binom{n-1+(n-j)2^{4k}+i2^{k}}{n-1}=2^{4k(n-1)}.
\end{equation}
\end{Claim}

\vspace{2mm}
\noindent
\textit{Proof of Claim~\ref{claim2}}
We prove this claim by induction on $n \ge 1$. It is easy to check that the statement is true for $n=1$.
Now, suppose that it is true for a fixed $n\ge 1$. 
It follows from the Pascal's relation that
\begin{eqnarray*}
\mathcal{L}
&:=&\sum_{j=0}^{n} (-1)^j\binom{n}{j}\binom{n+(n+1-j)2^{4k}+i2^k}{n} \\
&=& \sum_{j=0}^{n-1} (-1)^j\binom{n-1}{j}\binom{n+(n+1-j)2^{4k}+i2^k}{n} \\
&&- \sum_{j=0}^{n-1} (-1)^j\binom{n-1}{j}\binom{n+(n-j)2^{4k}+i2^k}{n}.
\end{eqnarray*}
Applying the Pascal's relation to the second binomial coefficient term in the first summation, 
we obtain from tedious but straightforward calculations that  
\begin{eqnarray*}
\mathcal{L}
&=&\sum_{j=0}^{n-1} (-1)^j\binom{n-1}{j} \binom{n-1+(n+1-j)2^{4k}+i2^k}{n} \\
&&+ \sum_{j=0}^{n-1} (-1)^j\binom{n-1}{j} \binom{n-1+(n+1-j)2^{4k}+i2^k}{n-1} \\
&&- \sum_{j=0}^{n-1} (-1)^j\binom{n-1}{j}\binom{n+(n-j)2^{4k}+i2^k}{n} \\
&=& \sum_{j=0}^{n-1} (-1)^j\binom{n-1}{j} \binom{n-1+(n+1-j)2^{4k}+i2^k}{n} \\
&&+ 2^{4k(n-1)}
- \sum_{j=0}^{n-1} (-1)^j\binom{n-1}{j}\binom{n+(n-j)2^{4k}+i2^k}{n} \\
&=& \sum_{j=0}^{n-1} (-1)^j\binom{n-1}{j} \binom{n-2+(n+1-j)2^{4k}+i2^k}{n} \\
&&+ 2\cdot2^{4k(n-1)}
- \sum_{j=0}^{n-1} (-1)^j\binom{n-1}{j}\binom{n+(n-j)2^{4k}+i2^k}{n},
\end{eqnarray*} 
where the second and the last equalities come from the induction hypothesis and Claim~\ref{claim1}.
Continuing this procedure $2^{4k}$ times, we can show that 
\begin{eqnarray*}
\mathcal{L}
&=& \sum_{j=0}^{n-1} (-1)^j\binom{n-1}{j}\binom{n+(n-j)2^{4k}+i2^k}{n} + 2^{4k} 2^{4k(n-1)} \\
&&- \sum_{j=0}^{n-1} (-1)^j\binom{n-1}{j}\binom{n+(n-j)2^{4k}+i2^k}{n} \\
&=& 2^{4kn},
\end{eqnarray*} 
which completes the proof.
\hfill$\square$

\vspace{2mm}
Now, we show that the number of $\tilde{y}$'s
satisfying $2^{3k}\tilde{y}_1+\cdots+ 2^{3k}\tilde{y}_n\equiv 0 \pmod  {2^{4k}}$ is
\begin{equation*}
\label{eq:sol}
\sum_{i=1}^{2^{3k}} \sum_{j=0}^{n-1}
(-1)^j\binom{n-1}{j}\binom{n-1+(n-j)2^{4k}+i2^k}{n-1},
\end{equation*}
which equals $2^{3k}2^{4k(n-1)}$, or equivalently, $2^{k(4n-1)}$
by Claim~\ref{claim2}.

Let us choose a sufficiently large number $q=4k$ such that $n < 2^k$, let 
$$
S_l=\{(\tilde{y}_1,\dots,\tilde{y}_n) \in \mathbb{Z}^n~|~ 
2^{3k}\tilde{y}_1+\cdots+ 2^{3k}\tilde{y}_n = l \cdot 2^{4k}, \tilde{y}_i \geq 0\}
$$
for each $l \in [n \cdot 2^{3k}-1]\cup \{0\}$, and let 
$$A_i^l=\{(\tilde{y}_1,\dots,\tilde{y}_n) \in S_l~| 
~\tilde{y}_i \geq 2^{4k} \} \subset S_l$$
for each $i=1,\dots,n$.
Then the number of integer solutions $\tilde{y} \in \mathbb{Z}_{2^{4k}}^n$ 
of the equation $2^{3k}\tilde{y}_1+\cdots+ 2^{3k}\tilde{y}_n\equiv 0 \pmod  {2^{4k}}$ is 
\[
\sum_{l=0}^{n \cdot 2^{3k}-1} \left| \left( S_l - A_1^l\right) \cap \cdots \cap \left( S_l - A_n^l \right)\right|
=\sum_{l=0}^{n \cdot 2^{3k}-1} \left|S_l - \left( A_1^l \cup\cdots \cup A_n^l \right)\right|,
\]
which can be exactly calculated as follows.

For each $t\cdot 2^{3k}\le l\le (t+1)\cdot 2^{3k}-1$ 
($t=0,1,\ldots,n-1$), 
by the inclusion-exclusion principle, we have 
\begin{eqnarray*}
\left|S_l - \left( A_1^l \cup\cdots \cup A_n^l \right)\right| 
&=& 
\left|S_l\right| - \left| A_1^l \cup\cdots \cup A_n^l\right| \\
&=& 
\binom{n-1+l2^k}{n-1}
+\sum_{I\subseteq[n]}
(-1)^{|I|}
\left|\bigcap_{j\in I}A_j^l\right|\\
&=& 
\sum_{j=0}^{t} (-1)^j
\binom{n}{j} \binom{n-1+l2^k-j2^{4k}}{n-1},
\end{eqnarray*}
where the last inequality is due to 
the fact that for $l<j2^{3k}$,
\[
\binom{n-1+l2^k-j2^{4k}}{n-1}=0.
\]

Let 
\begin{eqnarray*}
h(n)
&:=& 
\sum_{l=0}^{n \cdot 2^{3k}-1}
\left|S_l - \left( A_1^l \cup\cdots \cup A_n^l \right)\right| \\
&=&  
\sum_{t=0}^{n-1}
\sum_{i=0}^{2^{3k}-1}
\sum_{j=0}^{t} (-1)^j
\binom{n}{j} \binom{n-1+(t\cdot 2^{3k}+i)2^k-j2^{4k}}{n-1}.
\end{eqnarray*}
We now show that 
\begin{eqnarray*}
h(n)
&=&\sum_{i=0}^{2^{3k}-1} \sum_{j=0}^{n-1}
(-1)^j\binom{n-1}{j}\binom{n-1+(n-j)2^{4k}+i2^k}{n-1}\\
&=& 2^{k(4n-1)},
\end{eqnarray*}
where the second equality is due to Claim~\ref{claim2}.

Observe that
\begin{eqnarray*}
h(n)&=&
% \sum_{t=0}^{n}
% \sum_{i=0}^{2^{3k}-1}
% \sum_{j=0}^{t} (-1)^j
% \binom{n+1}{j} \binom{n+(t\cdot 2^{3k}+i)\cdot 2^k-j2^{4k}}{n} \\
% &=&
\sum_{t=0}^{n-1}
\sum_{i=0}^{2^{3k}-1}
\sum_{j=0}^{t} (-1)^j
\binom{n}{j} \binom{n-1+(t-j)2^{4k}+i2^k}{n-1} \\
&=& 
\sum_{t=0}^{n-1}
\sum_{i=0}^{2^{3k}-1}
\sum_{j=0}^{t} (-1)^j
\binom{n-1}{j} \binom{n-1+(t-j)2^{4k}+i2^k}{n-1} \\
&&+\sum_{t=1}^{n-1}
\sum_{i=0}^{2^{3k}-1}
\sum_{j=0}^{t} (-1)^j
\binom{n-1}{j-1} \binom{n-1+(t-j)2^{4k}+i2^k}{n-1}
\\
&=& 
\sum_{i=0}^{2^{3k}-1}
\sum_{j=0}^{n-1} (-1)^j
\binom{n-1}{j} \binom{n-1+(n-1-j)2^{4k}+i2^k}{n-1}
\\
&&+
\sum_{t=0}^{n-2}
\sum_{i=0}^{2^{3k}-1}
\sum_{j=0}^{t} (-1)^j
\binom{n-1}{j} \binom{n-1+(t-j)2^{4k}+i2^k}{n-1} \\
&&+\sum_{t=1}^{n-1}
\sum_{i=0}^{2^{3k}-1}
\sum_{j=0}^{t} (-1)^j
\binom{n-1}{j-1} \binom{n-1+(t-j)2^{4k}+i2^k}{n-1},
\end{eqnarray*}
where the second equality comes from the Pascal's relation. 
It follows from Claim~\ref{claim1} (or Eq.~ (\ref{eq:claim11})) 
and Claim~\ref{claim2} that 
\begin{eqnarray*}
h(n)
&=& 
\sum_{i=0}^{2^{3k}-1}
\sum_{j=0}^{n-1} (-1)^j
\binom{n-1}{j} \binom{n-1+(n-j)2^{4k}+i2^k}{n-1}
\\
&&+
\sum_{t=0}^{n-2}
\sum_{i=0}^{2^{3k}-1}
\sum_{j=0}^{t} (-1)^j
\binom{n-1}{j} \binom{n-1+(t-j)2^{4k}+i2^k}{n-1} \\
&&+\sum_{t=1}^{n-1}
\sum_{i=0}^{2^{3k}-1}
\sum_{j=0}^{t} (-1)^j
\binom{n-1}{j-1} \binom{n-1+(t-j)2^{4k}+i2^k}{n-1}\\
&=& 
2^{k({4n}-1)}
\\
&&+
\sum_{t=0}^{n-2}
\sum_{i=0}^{2^{3k}-1}
\sum_{j=0}^{t} (-1)^j
\binom{n-1}{j} \binom{n-1+(t-j)2^{4k}+i2^k}{n-1} \\
&&+\sum_{t=0}^{n-2}
\sum_{i=0}^{2^{3k}-1}
\sum_{j=1}^{t+1} (-1)^j
\binom{n-1}{j-1} \binom{n-1+(t+1-j)2^{4k}+i2^k}{n-1}\\
&=& 
2^{k({4n}-1)}
\\
&&+
\sum_{t=0}^{n-2}
\sum_{i=0}^{2^{3k}-1}
\sum_{j=0}^{t} (-1)^j
\binom{n-1}{j} \binom{n-1+(t-j)2^{4k}+i2^k}{n-1} \\
&&+\sum_{t=0}^{n-2}
\sum_{i=0}^{2^{3k}-1}
\sum_{j=0}^{t} (-1)^{j+1}
\binom{n-1}{j} \binom{n-1+(t+1-j-1)2^{4k}+i2^k}{n-1}\\
&=& 2^{k({4n}-1)}.
\end{eqnarray*}

Since the number of all possible $\tilde{y}$ is $2^{4kn}$, 
the probability that $\tilde{y} \in \mathbb{Z}_{2^{4k}}^n$ is orthogonal to 
$\tilde{u}$ is at most $2^{k(4n-1)}/2^{4kn}$ which is equal to $1/2^{k}$. 

\end{document}